\input{aipcheck}

\documentclass[
    ,final            % use final for the camera ready runs
  ]
  {aipproc}

\usepackage[dvips]{color}
\usepackage{amsmath}
\usepackage{amsfonts}
\usepackage{amssymb}

\usepackage{graphics}
\usepackage{amsmath}
\usepackage{amsfonts}
\usepackage{amssymb}
\usepackage{graphicx}
\usepackage{epsfig}

\def\ba{\begin{eqnarray}}
\def\ea{\end{eqnarray}}
\def\be{\begin{equation}}
\def\ee{\end{equation}}

\def\d{\mathrm{d}}

\def\mn{_{\mu \nu}}

\def\({\left(}
\def\){\right)}

\def\ie{{\it i.e. }}

\definecolor{dunkelblau}{rgb}{0, 0, .7}
\definecolor{dunkelmagenta}{rgb}{1, 0.2, 1}
\definecolor{dunkelmagenta}{rgb}{0.95, 0.2, 0.8}
\definecolor{dunkelmagenta}{rgb}{1, 0, 0}
\definecolor{cyan}{rgb}{0,0,0.7}
\definecolor{cyand}{rgb}{0,0,0.7}

\layoutstyle{6x9}

\begin{document}

\title{Cascading Gravity and Degravitation}

\classification{11.10.Gh, 11.10.Kk, 11.25.-w}
\keywords      {dark energy, massive gravity, large extra dimensions, six-dimensional gravity}

\author{Claudia de Rham}{
address={Dept. of Physics \& Astronomy, McMaster University, Hamilton ON, Canada\\
Perimeter Institute for Theoretical Physics,  Waterloo, ON, Canada}
}
%
%\author{<author2>}{
%  address={<common address for author2 and author3>}
%}
%
%\author{<author3>}{
%  address={<common address for author2 and author3>}
%  ,altaddress={<author1 address>} % additional visiting address
%}

\begin{abstract}
Cascading gravity is an explicit realization of the idea of degravitation,
where gravity behaves as a high-pass filter. This could explain why a
large cosmological constant does not backreact as much as
anticipated from standard General Relativity. The model relies on the
presence of at least two infinite extra dimensions while our world
is confined on a four-dimensional brane. Gravity is then four-dimensional
at short distances and becomes weaker and weaker at larger
distances.

%The curvature on codimension-two and higher branes is not
%regular for arbitrary matter sources. Nevertheless, the low-energy theory
%for an observer on such a brane should be well-defined and
%independent to any regularization procedure. This is achieved via
%appropriate {\it classical} renormalization of the brane
%couplings, and leads to a natural hierarchy between standard model
%couplings and couplings to gravity.
\end{abstract}

\maketitle

%%%%%%%%%%%%%%%%%%%%%%%%%%%%%%%%%%%%%%%%%%%%
%% MAINMATTER
%%%%%%%%%%%%%%%%%%%%%%%%%%%%%%%%%%%%%%%%%%%%

\subsection{Degravitation and massive gravity}

The cosmological constant problem is one of the most tantalizing
puzzles of modern cosmology. While most approaches propose an
explanation for what dark energy is,
and why the Universe is accelerating, very little progress has been
made in understanding why the observed cosmological constant is so
much smaller than expected in the first place.
Instead, we follow here an alternative approach based on the idea of
degravitation, see ref.~\cite{Dvali:2007kt}, where the cosmological
constant could be as large as expected from standard field theory,
but would simply gravitate very little. In models
where the graviton has a mass, gravity is typically weaker in the
infrared (IR) and could play the role of a high-pass filter.

The idea behind degravitation is to promote the mass of the graviton
in the Fierz-Pauli action to an operator $m^2(\Box)=\(L^2 \Box\)^\alpha/L^2$, such that the
inverse propagator is then of the form
\ba
\label{alphaProp}
\mathcal{G}^{-1}=\Box\(1-\frac{m^2(\Box)}{\Box}\)=\Box\(1-\(L^2
\Box\)^{\alpha-1}\)\,.
\ea
To satisfy unitarity, the power $\alpha$ should be positive, while to
represent an IR modification of gravity one should have $\alpha<1$.
One can further show that in order for the resulting theory to play
the role of a high-pass filter, the power should satisfy $\alpha
<1/2$,~\cite{Dvali:2007kt,Dvali:2006}.
The extra-dimensional DGP scenario proposed by
Dvali, Gabadadze and Porrati, see ref.~\cite{Dvali:2000rv}, represents a
specific model of massive gravity, with the parameter
$\alpha=1/2$. Lying precisely on the border, this model does
not represent an explicit non-linear realization of
degravitation but is worth studying as it provides valuable
intuition. We will then extend the DGP scenario to higher
codimensions and present the first non-linear realization
of a high-pass filter for a spin-2 particle.

\subsection{DGP as an example of massive gravity}

The DGP scenario relies on the idea that our four-dimensional world
could be a hypersurface or brane embedded in a five-dimensional
space-time. If the volume of the extra dimension were finite, the
effective four-dimensional theory would admit a separate zero-mode
that would dominate in the IR. In order to represent a modification
of gravity in the IR,
it is therefore imperative to keep the volume of
the extra dimension infinite. Without any further ingredient,
gravity would then be five-dimensional at all scales, and clearly
incompatible with Newton's law. To recover four-dimensional
gravity at short scales, the DGP model relies on
a four-dimensional Einstein-Hilbert term localized on the brane,
leading to the action
\ba
S_{\rm{DGP}}=\int \d ^5x \sqrt{-g_5}\ \frac{M_5^3}{2}\ R^{(5)}
+\int \d^4x \sqrt{-g_4}\(\frac{M_4^2}{2}R^{(4)}+\mathcal{L}_m\)\,,
\ea
where $M_n$ is the $n$-dimensional Planck mass, $R^{(n)}$ the scalar
curvature with respect to the $n$-dimensional metric, and
$\mathcal{L}_m$ the Lagrangian of matter fields localized on the
brane. In five-dimensions, gravity is propagated by a spin-2 massless
field which has five degrees of freedom. From a
four-dimensional point of view, on the other hand, these five degrees of freedom will
play the role of a massive spin-2 particle, and the effective
theory on the brane will be an explicit non-linear realization of
massive gravity. The effective four-dimensional propagator in this
model is then
\ba
\mathcal{G}(k)=\frac{1}{M_4^2}\frac{1}{k^2+m_5k}\,
\ea
where $k$ is the four-dimensional momentum and the ``graviton mass'' is
$m_5=M_5^3/M_4^2$. This propagator is precisely of the form
presented in \eqref{alphaProp}, with the parameter $\alpha=1/2$. At
short distances compared to $1/m_5$, the four-dimensional curvature
terms dominate and gravity appears four-dimensional, with the
standard $1/r^2$ Newton's law. At larger distances, on the other
hand, gravity probes the extra dimension and gravity behaves fully
five-dimensional, with force law going as $1/r^3$, gravity is
therefore successfully modified in the IR. As mentioned in the
previous section, this modification of gravity is however not
sufficient
To represent an explicit realization of a high-pass filter, and we now
turn to a higher dimensional generalization of the DGP model.

\subsection{Higher codimension sources}

In a $(4+n)$-dimensional space-time, the gravitational potential
goes as $V(r)\sim r^{-(1+n)}$. In the spectral representation of the
Newtonian propagator
\ba
V(r)=\int_0^\infty \frac{\rho(m^2)e^{-m r}}{r}\ \d m^2\,,
\ea
this corresponds to a spectral density $\rho(m^2)\sim m^{n-2}$ as $m\to
0$.
%
%and the four-dimensional propagator
%on a four-dimensional brane goes therefore as $\mathcal{G}(k)\sim
%k^{n-2}$. If dominated by the mass term $m^2(k)\sim k^{2\alpha}$.
The propagator then has a spectral representation
\ba
\mathcal{G}(k)=\int_0^\infty \frac{\rho(m^2)}{k^2+m^2}\ \d m^2\,.
\ea
For $n=1$ extra dimensions, as in DGP, this gives a propagator of the form $\mathcal{G}(k)\sim
1/k$ in the IR limit and therefore corresponds to a theory with $\alpha=1/2$.
For any higher dimensional model, $n\ge2$, the propagator tends to a
constant in the deep IR limit $k\to 0$ and therefore corresponds to
a theory of massive gravity with $\alpha=0$ which exhibits the
right behaviour for degravitation which needs $0\le \alpha<1/2$.

When dealing with more than one extra dimension, great care should
however be taken in dealing with divergences that arise on
objects of codimension greater or equal to two, see
ref.~\cite{Geroch:1987qn,Goldberger:2001tn}. The only source on a codimension-two
object (such as a cosmic string) that does not give rise to
curvature divergences is a pure tension. Any other kind of matter
leads to logarithmic divergences on the defect. To avoid such
divergences, the defect should first be regularized.

\subsection{Scalar field toy-model}

To see this, one can study the propagator in a scalar field toy-model.
If the scalar field is sourced on a codimension-one,
\ba
\partial_y^2 \phi(y)=\delta(y)
\ea
the field satisfies $\phi(y)\sim |y|$ and remains regular on the
object at $y=0$. If instead the source is localized on a codimension-two
surface,
\ba
\(\partial_y^2+\partial_z^2\)\phi(y,z)=\delta^2(y,z)
\ea
the field configuration is then $\phi(y,z)\sim \log(y^2+z^2)$ and
hence diverges logarithmically as it approaches the defect $y,z\to
0$.
However if we instead embedded the codimension-two object on a
codimension-one brane on where a field kinetic term is included,
\ba
\(\partial_y^2+\partial_z^2\)\phi(y,z)=\delta^2(y,z)-\delta(z)\,
\partial_y^2 \phi\,,
\ea
the field would then be regular on the codimension-two object, \cite{deRham:2007rw,deRham:2007xp}.
%The
%new kinetic term on the codimension-one plays the role of a
%regulator, .
In a six-dimensional setup, this theory follows from the action
\ba
S=-\int \d^6x \frac{M_6^4}{2} \partial_A \phi\, \partial^A \phi
-\int \d^5x \frac{M_5^3}{2} \partial_\alpha \phi\, \partial^\alpha \phi
-\int \d^4x \frac{M_4^2}{2} \partial_\mu \phi\, \partial^\mu \phi\,,
\ea
which corresponds to the effective four-dimensional propagator
\ba
\mathcal{G}_{4d}=\frac{1}{M_4^2}\frac{1}{k^2+m^2(k^2)}\,,\hspace{15pt}{\rm{with}}\hspace{15pt}
m^2(k^2)=\frac{\pi
m_5}{4}\frac{\sqrt{m_6^2-k^2}}{{\rm{arcth}}\(\sqrt{\frac{m_6-k}{m_6+k}}\)}\,,
\ea
and $m_6=M_6^4/M_5^3$. From a four-dimensional point of view, this
corresponds to a theory of massive gravity with $\alpha=0$. At long wavelengths,
$k\ll m_6$ the propagator behaves as in six dimensions, while for $m_6\ll k\ll
m_5$, the field effectively sees five dimensions and at shorter
wavelengths $k\gg m_5$ one recovers a four-dimensional propagator.
This behavior has also been obtained in a different codimension-two
framework, \cite{Kaloper:2007ap}.

This cascading approach that avoids any singularities when dealing
with higher codimensions is completely generalizable to any number
of dimensions. From a four-dimensional point of view, gravity will
then go from four dimensions at small scales, up to five dimensions,
etc. up to $(n+4)$ dimensions at large scales, exhibiting a theory
of massive gravity with $\alpha=0$.

\subsection{Cascading Gravity}

The generalization to gravity is straightforward,
\ba
S=-\int \d^6x \sqrt{-g_{6}}\ \frac{M_6^4}{2} R^{(6)}
-\int \d^5x \sqrt{-g_{5}}\ \frac{M_5^3}{2} R^{(5)}
-\int \d^4x \sqrt{-g_{4}}\ \frac{M_4^2}{2} R^{(4)}\,.
\ea
To study the gravitational propagator, one can work around flat space-time, $g_{AB}=\eta_{AB}+h_{AB}$,
and decompose the four-dimensional perturbations as $h\mn=h\mn^{TT}+\pi \, \eta\mn+$gauge terms.
The tensor mode $h\mn^{TT}$ behaves precisely as the scalar field toy-model, and
is regularized by the presence of the five-dimensional
Einstein-Hilbert term. The scalar mode $\pi$, on the other hand, propagates a
ghost. After integrating the sixth dimension, its effective equation
on the five-dimensional brane is
\ba
-\frac{M_5^3}{2}\left[\Box_4-m_6 \sqrt{-\Box_4}\right]\pi
=\frac{1}{12}\(T^{(4)}-3M_4^2\Box_4\pi\)\,.
\ea
$\pi$ is therefore finite on the codimension-two brane.
However in the UV, $\pi\sim +\Box_{4}^{-1}T$ while in the IR, $\pi\sim - T$.
The kinetic term hence changes sign signaling the presence of a
ghost. This ghost is completely independent to the ghost present in
the self-accelerating branch of DGP but is completely generic to any
codimension-two and higher framework with brane localized kinetic
terms. In particular this ghost is also present when considering a
pure codimension-two scenario with no five-dimensional
Einstein-Hilbert term, \cite{Gabadadze:2003ck}.

There are however two ways to cure the ghost, both of which are natural when considering a
realistic higher codimensional scenario.
First of all, we do expect a large cosmological constant on the
brane, and in the presence of a minimal tension on the brane, the
ghost disappears and is replaced by a healthy mode,  \cite{deRham:2007rw}.
Therefore this provides a realistic resolution to the ghost issue
and shows that this model of cascading gravity with
a brane tension is free of any instability, at least around flat space-time.

Secondly, one never thinks of any defect as being fundamentally
thin. Any distributional source, is well described by a delta
function at low-energy, but there is nonetheless an energy scale
above which the thickness of the brane is resolved, and the brane
can no longer be treated as a thin object. This is the case in this
cascading scenario, and another independent way to cure the ghost is
to smooth the brane, \cite{deRham:2007xp}. In the underlying theory, each localized
Einstein-Hilbert term is nothing but a fundamental six-dimensional
scalar curvature object localized on the thick brane. One can check
that the ghost disappears and is replaced by a healthy mode when
properly taking this into account before taking the thin-brane limit
to derive the low-energy effective theory, \cite{deRham:2007xp,Gabadadze:2003ck}.

When properly taking into account this issue associated with the
ghost, we recover a theory of massive gravity composed of one
helicity-2 mode, helicity-one modes that decouple and 2 helicity-0
modes. In order for this theory to be consistent with standard
general relativity in four dimensions, both helicity-0 modes should
decouple from the theory. This decoupling does not happen in a
trivial way, and relies on a phenomena of strong coupling. Close
enough to any source, both scalar modes are strongly coupled and therefore freeze.

The resulting theory appears as a theory of a massless spin-2 field in four-dimensions,
in other words as General Relativity.
If $r\ll m_5$ and for $m_6\le m_5$, the respective Vainshtein scale or strong coupling scale, \ie the distance from the source $M$
within which each mode is strongly coupled is $r_{i}^3=M/m_i^2
M_4^2$, where $i=5,6$. Around a source $M$, one recovers
four-dimensional gravity for $r\ll r_{5}$, five-dimensional
gravity for $r_{5}\ll r \ll r_{6}$ and finally six-dimensional
gravity at larger distances $r\gg r_{6}$.

\subsection{Discussion}

Models of massive gravity represent a novel framework for tackling the
cosmological constant problem. There is to date only one
known ghost-free non-linear realization of massive gravity that does
not violate Lorenz invariance, thats is the DGP model (perturbed around its stable
branch), and its higher-codimensional generalization: \ie
cascading gravity. Since the DGP model does not exhibit the phenomena of degravitation,
cascading gravity represents the only non-linearized
explicit candidate for degravitation known so far. Such a theory
evades most tests of General Relativity due to strong coupling and
is hence compatible with solar systems tests, but represents
nonetheless a high-pass filter theory of gravity and slowly filters
out in time sources with long wavelength sources such as the cosmological
constant.

%%%%%%%%%%%%%%%%%%%%%%%%%%%%%%%%%%%%%%%%%%%%%%%%
%% BACKMATTER
%%%%%%%%%%%%%%%%%%%%%%%%%%%%%%%%%%%%%%%%%%%%%%%%

\begin{theacknowledgments}
This talk is based on work in collaboration with Gia Dvali, Justin Khoury,
Stefan Hofmann, Oriol Pujol\`as, Michele Redi and Andrew J. Tolley.
Research at McMaster is supported by the Natural
Sciences and Engineering Research Council of Canada.
Research at Perimeter Institute for Theoretical Physics is supported
in part by the Government of Canada through NSERC and by the
Province of Ontario through MRI.
\end{theacknowledgments}

\bibliographystyle{aipproc}
\IfFileExists{\jobname.bbl}{}
 {\typeout{}
  \typeout{******************************************}
  \typeout{** Please run "bibtex \jobname" to optain}
  \typeout{** the bibliography and then re-run LaTeX}
  \typeout{** twice to fix the references!}
  \typeout{******************************************}
  \typeout{}
 }

\end{document}